\documentclass[iop]{emulateapj}

\usepackage{longtable}
\usepackage{rotating}
\def\simlt{\lower.5ex\hbox{$\; \buildrel < \over \sim \;$}}

\begin{document}

\shorttitle{The MCR\ \&\ SHMR in CS82}
\shortauthors{Shan et al.}
\title{The Mass-Concentration Relation and the Stellar-to-Halo Mass Ratio in the CFHT Stripe 82 Survey}
\author{HuanYuan Shan\altaffilmark{$\dagger$1,2}
Jean-Paul Kneib\altaffilmark{1,3},
Ran Li\altaffilmark{4},
Johan Comparat\altaffilmark{5,6},
Thomas Erben\altaffilmark{2},
Martin Makler\altaffilmark{7},
Bruno Moraes\altaffilmark{8,9},
Ludovic Van Waerbeke\altaffilmark{10},
James E. Taylor\altaffilmark{11},
Ald\'ee Charbonnier\altaffilmark{12},
Maria E. S. Pereira\altaffilmark{7}}
\altaffiltext{1}{Laboratoire d'astrophysique (LASTRO), Ecole Polytechnique F\'ed\'erale de Lausanne (EPFL), Observatoire de Sauverny, CH-1290 Versoix, Switzerland}
\altaffiltext{2}{Argelander Institute for Astronomy, University of Bonn, Auf dem H{\"u}gel 71, 53121 Bonn, Germany}
\altaffiltext{3}{Aix Marseille Universit\'e, CNRS, LAM (Laboratoire d'Astrophysique de Marseille) UMR 7326, 13388, Marseille, France}
\altaffiltext{4}{Key Laboratory for Computational Astrophysics, The Partner Group of Max Planck Institute for Astrophysics, 
National Astronomy Observatory, Chinese Academy of Sciences, Beijing 100871, China}
\altaffiltext{5}{Departamento de Fisica Teorica, Universidad Autonoma de Madrid, Spain}
\altaffiltext{6}{Instituto de Fisica Teorica UAM/CSIC, Spain}
\altaffiltext{7}{Centro Brasileiro de Pesquisas F\'isicas, Rua Dr Xavier Sigaud 150, CEP 22290-180, Rio de Janeiro, RJ, Brazil}
\altaffiltext{8}{Department of Physics and Astronomy, University College London, Gower Street, London, WC1E 6BT, UK}
\altaffiltext{9}{CAPES Foundation, Ministry of Education of Brazil,  Brasilia/DF 70040-020, Brazil}

\altaffiltext{10}{Department of Physics and Astronomy, University of British Columbia, 6224 Agricultural Road, Vancouver, V6T 1Z1, BC, Canada}
\altaffiltext{11}{Department of Physics and Astronomy, University of Waterloo, 200 University Avenue West, Waterloo, Ontario, Canada N2L 3G1}
\altaffiltext{12}{Observat\'orio do Valongo, Universidade Federal do Rio de Janeiro, Ladeira do Pedro Ant\^onio 43, Sa\'ude, Rio de Janeiro, RJ 20080-090, 
Brazil \& Centro Brasileiro de Pesquisas F\'isicas, Rua Dr. Xavier Sigaud 150, Rio de Janeiro, RJ 22290-180, Brazil
}
   \altaffiltext{$\dagger$}{Email address: {\tt shanhuany@gmail.com}}

\begin{abstract}
We present a new measurement of the mass-concentration relation and the stellar-to-halo mass ratio over the 
halo mass range $5\times 10^{12}$ to $2\times 10^{14}M_{\odot}$. To achieve this, we use weak lensing 
measurements from the  CFHT Stripe 82 Survey (CS82), combined with the central galaxies from the redMaPPer cluster catalogue 
and the LOWZ/CMASS galaxy sample of the Sloan Digital Sky Survey-III Baryon Oscillation Spectroscopic Survey Tenth Data 
Release. The stacked lensing signals around these samples are modeled as a sum of 
contributions from the central galaxy, its dark matter halo, and the neighboring halos, as well as a term for possible centering errors. We measure 
the mass-concentration relation: $c_{200c}(M)=A(\frac{M_{200c}}{M_0})^{B}$ with 
$A=5.24\pm1.24, B=-0.13\pm0.10$ for $0.2<z<0.4$ and $A=6.61\pm0.75, B=-0.15\pm0.05$ for $0.4<z<0.6$. 
These amplitudes and slopes are completely consistent with predictions from recent simulations. 
We also measure the stellar-to-halo mass ratio for our samples, and find results consistent with previous measurements 
from lensing and other techniques.
\end{abstract}


\section{Introduction}

Dark Matter (DM) is the dominant mass component in the universe. The hierarchical cold DM model with a cosmological constant ($\Lambda$CDM) is successful in explaining many observations of large-scale structure, and the growth and evolution of this structure with redshift. On smaller scales, simulations of structure formation in the DM component predict that matter should cluster in a characteristic way, producing dense, virialized halos with a universal density profile. The original description of this profile, by Navarro, Frenk and White (NFW -- Navarro et al. 1997), is characterized by just two parameters: these can be taken to be, for instance, the radius of the virialized region $r_{vir}$ and the `scale' radius $r_s$ at which the density profile has a logarithmic slope $d\ln \rho/d \ln r = -2$. Equivalently, halos  can be described by a virialized mass (the mean density within the virial radius being fixed  at a given redshift) and a concentration parameter $c$, the ratio of the virial radius to the scale radius. Simulations have shown that these two parameters are correlated, with the average concentration of a halo being a weakly decreasing function of mass for most halos (e.g. NFW; Jing 2000; Bullock et al. 2001.)
 
More recent work at higher resolution has shown that the density profiles of the most massive halos in particular deviate slightly from the original profile proposed by NFW, and are better fit by an Einasto profile (Einasto 1965; cf.~Merritt et al. 2006; Gao et al. 2008; Dutton \& Maccio 2014; Klypin et al. 2014), but for most systems the  difference between the fits is relatively minor, and the behaviour of the concentration parameter, taken as the ratio of the virial radius to the radius where $d\ln \rho/d \ln r = -2$, is approximately unchanged 
(though see Klypin et al.~2014 for a generalized definition of concentration, with a slightly different behaviour at high mass.)

The concentration parameter appears to trace the formation history of halos, and the mean mass-concentration (\textit{m-c} hereafter) relation is cosmology dependent (e.g.~Zhao et al. 2003; 2009). Thus it represents an interesting cosmological measurement for surveys of groups and clusters. Individual masses and concentrations, and/or the average \textit{m-c} relation, have been measured using several different methods, including gravitational lensing, X-ray surface brightness profile fitting, and galaxy velocities (see Groener et al. 2016 for references and a summary of techniques). Lensing-based methods are particularly interesting because they can determine the total mass distribution, without making any assumptions about the baryonic contents of a halo; on the other hand they measure projected density, and need to be interpreted carefully as a result. Specific lensing techniques for estimating concentration or the density profile include strong gravitational lensing (e.g. Comerford \& Natarajan 2007) weak gravitational lensing (using shear around stacked samples, e.g.~Mandelbaum et al.~2008; Fo\"{e}x et al. 2014; shear around single massive clusters, e.g.~Okabe et al. 2010; Du \& Fan 2014; or shear + magnification, e.g.~Umetsu et al.~2014), and combinations of weak and strong lensing (Oguri et al. 2009, 2012; Sereno \& Covone 2013; Merten et al. 2014).

Although current results from large samples are in reasonable agreement with theoretical predictions of the \textit{m-c} relation (e.g.~Merten et al. 2014; Umetsu et al.~2014), several concerns remain. The first is that many cluster samples are selected for x-ray luminosity and morphology, and/or the presence of strong lensing. Either selection can bias the sample to more concentrated objects, and/or prolate systems elongated along the line of sight (e.g.~Oguri et al. 2005; Hennawi et al. 2007; Meneghetti et al. 2014). Clearly stacked measurements over large, volume-limited samples of optically selected clusters have an advantage here. A second concern is that the slope of the \textit{m-c} relation is expected to be very shallow (e.g.~Klypin et al.~2014), so measurements at low masses are particularly useful to constrain it. Here too, large samples are required to obtain a sufficiently strong stacked lensing signal. For both these reasons, lensing data from a deep, large-area survey is particularly useful for constraining halo concentration.

Stellar-mass-selected samples with well calibrated mean halo masses have several other uses.
The luminosity or stellar mass of the principal galaxy in a system should correlate strongly with the properties of the halo surrounding it.
The exact relationship between central stellar mass and DM halo mass, and the scatter in this relationship, are critical for understanding the formation and evolution of galaxies. The stellar mass of the central galaxy in a system can be inferred from its observed luminosity in one or more bands. There are then several methods for estimating the mass of the surrounding halo, including stellar kinematics (Conroy et al. 2007; More et al. 2011; Wojtak \& Mamon 2013), abundance matching (Moster et al. 2010; Behroozi et al. 2010; Shankar et al. 2014), the Tully-Fisher relation (Pizagno et al. 2007; Miller et al. 2014), weak lensing (Mandelbaum et al. 2006; Velander et al. 2014; Hudson et al. 2015; Han et al. 2015), and combinations of weak lensing and clustering (Leauthaud et al. 2012). Comparing the two, one can derive the mean stellar-to-halo mass relation (\textit{SHMR}) or its inverse, the mean halo mass at fixed stellar mass.

Collectively, these studies have found a \textit{SHMR} that is close to a broken power law, with stellar mass increasing rapidly with halo mass for low-mass systems, and more slowly for high-mass systems. The break in the relationship occurs at $M_*\gtrsim 4.5\times 10^{10}M_{\odot}$, or $M_h \sim 1.5\times 10^{12} M_{\odot}$ (e.g.~Leauthaud et al. 2012), around the transition from isolated galaxies to galaxy groups. Given the importance of this measurement in relating dark matter structure to visible galaxies, it should be confirmed and calibrated independently in multiple surveys, with different potential selection functions and systematics. Once again, large halo samples and deep lensing measurements are required to get accurate results.

The CFHT Stripe 82 survey is a deeper $i$-band follow up of the SDSS Stripe 82 region with MegaCam on CFHT. Relative to SDSS imaging, it provides increased depth and excellent seeing, allowing sensitive lensing measurements over more than 100 deg$^2$. Thus it represents an ideal data set to explore the scaling of halo properties with mass. In this paper, we use a sample of redMaPPer clusters (Rykoff et al. 2014) and the LOWZ and CMASS galaxy samples of SDSS-III BOSS DR10 (Ahn et al. 2014), together with the shear catalog from the CFHT Stripe 82 dataset to measure the tangential shear around central galaxies, and to quantify the \textit{m-c} relation and the \textit{SHMR}. Throughout the paper, we assume a $\Lambda$CDM cosmology with parameter values from the latest Planck Collaboration analysis (Ade et al. 2014): $\Omega_m=0.315$, $\Omega_{\Lambda}=0.685$, $\sigma_8=0.829$, $n_s=0.96$, and
$H_0=100~h~\mathrm{km}~{\mathrm s}^{-1}~\mathrm{Mpc}^{-1}$ with $h=0.673$. Magnitudes are expressed using the AB scale.

\section{OBSERVATIONAL DATA} \label{sec: data}

\subsection{The Source catalog}

The source galaxies used in our measurement are taken from the CFHT Stripe 82 
Survey (CS82: Erben et al. 2015), which is an $i$-band imaging survey covering a large fraction of SDSS Stripe 82 
region with a median seeing $0.59^{\prime\prime}$. The survey contains a total of $173$ tiles with $165$ from CS82 and $8$ 
from CFHT-LS Wide. Each CS82 tile was obtained from $4$ consecutive dithered observations, each with an exposure time
of $410$ seconds. The completeness magnitude is $i_{AB}\sim24.0$.

We use all galaxies with weight $w>0$, FITCLASS$=$0 and MASK$\leq$1, in which $w$ represents an inverse variance 
weight assigned to each source galaxy by \textsc{lensfit}, while FITSCLASS is a star/galaxy classification provided by 
\textsc{lensfit}. The parameter MASK describes the mask information at an object's position. Objects with MASK$\leq$1 can 
safely be used for most weak lensing analyses (Erben et al. 2013). After masking out
bright stars and other image artifacts, the effective sky coverage is $\sim$130$\rm deg^2$.

The shapes of faint galaxies are measured with the \textsc{lensfit} method (Miller et al. 2007, 2013). 
In this work we use the same pipeline as the CFHTLenS collaboration. The \textsc{lensfit} algorithm, as applied to 
CFHTLenS data, was calibrated by Miller et al. (2013) 
using two sets of simulated images, the GREAT CFHTLensS Image Simulation and 
the SkyMaker CFHTLenS Image Simulation, covering a wide range of different observing conditions 
and PSFs. The CFHTLenS data itself was also taken over a wide range of seeing, number of exposures, noise 
and depth. Miller et al. (2013) found that the shape measurement errors, including the multiplicative bias factor $m$ and the 
additive bias $c$, can be well modelled as a function of galaxy signal-to-noise ratio and size over this entire range 
of conditions. Thus, although the mean observing conditions of CS82 are different from those of CFHTLenS, the 
\textsc{lensfit} pipeline and calibration can also be applied to CS82 data.

As CS82 contains only $i$-band imaging, we derive photometric redshifts (photo-zs) for the source galaxies from
the overlapping {\textit{ugriz}} co-add data from SDSS (Annis et al. 2014). Photo-zs are derived using the
code \textsc{BPZ} (Ben{\'{\i}}tez 2000), following the method described in Bundy et al. (2015). The resulting 
catalogue is the same as was used in Battaglia et al. (2015), and cuts and systematics tests of the catalogue 
are described in detail there. Briefly, the lensing catalogue is limited to objects with photo-zs, corresponding 
to objects with reliable detections in the SDSS co-add. To limit the number of catastrophic redshift errors, 
objects with a \textsc{BPZ} odds parameter less than $0.5$ are also cut from the catalogue. After these cuts, 
the galaxy number density for the source catalog is roughly $4.5~\rm gals/arcmin^2$.

Li et al. (2016) have compared the lensing signal of CMASS and LOWZ galaxies with source galaxies from different redshift bins and shown that the results agree with each within statistical error bars.

\subsection{Lens selection}

We consider as foreground lenses central galaxies from the redMaPPer cluster catalogue, as well as individual massive galaxies from the LOWZ/CMASS samples. We separate the lenses into several bins in stellar mass, and two redshift bins, $0.2<z<0.4$ and $0.4<z<0.6$.

\subsubsection{The RedMaPPer Cluster Catalog} 

The redMaPPer cluster catalog by Rykoff et al. (2014) is constructed from photometric galaxy samples of the SDSS DR8, 
using the optimized red-sequence richness estimator $\lambda$. The richness parameter $\lambda$ corresponds to the number of red 
sequence galaxies brighter than $0.2L_{*}$ at the redshift of the cluster, within a scaled aperture. 
In order to reduce the potential effects of miscentering, we choose only clusters with well-defined centers, corresponding to centering probabilities $\rm P_{\rm cen}>0.9$ (see Rykoff et al.~2014).  We then divide the clusters into two redshift bins with redshift $0.2<z<0.4$ and $0.4<z<0.6$, and two richness bins ($20<\lambda<30, 30<\lambda<200$) at each redshift. The final samples contain ($14/118$) systems at high redshift and ($134/124$) systems at low redshift, respectively. 

\subsubsection{LOWZ \& CMASS Galaxy Catalog}

We also use the LOWZ and CMASS galaxy sample from SDSS-III BOSS DR10 as lens 
galaxies\footnote{\tt https://www.sdss3.org/dr10/spectro/galaxy.php}. 
The BOSS survey has measured spectroscopic redshifts for $1.5$ million galaxies, and is approximately volume limited to $z\sim0.52$. The two samples are both the main SDSS-III BOSS BAO tracers (Dawson et al. 2013). 
The LOWZ samples consist of red galaxies at $z<0.4$ from the SDSS DR8 (Aihara et al. 2011) imaging data; the CMASS samples are selected with an approximately constant stellar mass threshold (Eisenstein et al. 2011). 
Stellar masses for the two samples were derived using the Portsmouth SED-fitting method (Maraston et al. 2013), based on the BOSS 
spectroscopic redshift. In calculating the stellar mass, the Kroupa (2001) initial mass function (IMF) was assumed.

We choose LOWZ and CMASS galaxies in our two redshift ranges, $0.2<z<0.4$ and $0.4<z<0.6$. 
We also divide them into $3$ and $4$ stellar mass bins, which will be used on the \textit{SHMR} measurement, 
with ($2350, 2492, 2325$) and ($5966, 6168, 7481, 7611$) systems for the low-z and high-z range, respectively (see Figure~1).

\section{Lensing Signal}

We measure the \textit{m-c} relation using the stacked lensing signal. Stacking many 
halos reduces the fluctuations due to noise caused by uncorrelated structures along the line of sight, 
shape measurement noise, substructures, and the shape variations of individual halos. Thus, the measurement 
can more accurately determine the average mass profile. Furthermore, it allows for the lensing measurement of 
lower mass halos, where individual detection is impossible due to their smaller shears relative to clusters. 
Individual massive cluster observations and those based on stacked analysis of many halos are thus 
complementary, drastically increasing the available sensitivity in mass.

\begin{figure*}
\includegraphics[width=0.49\textwidth]{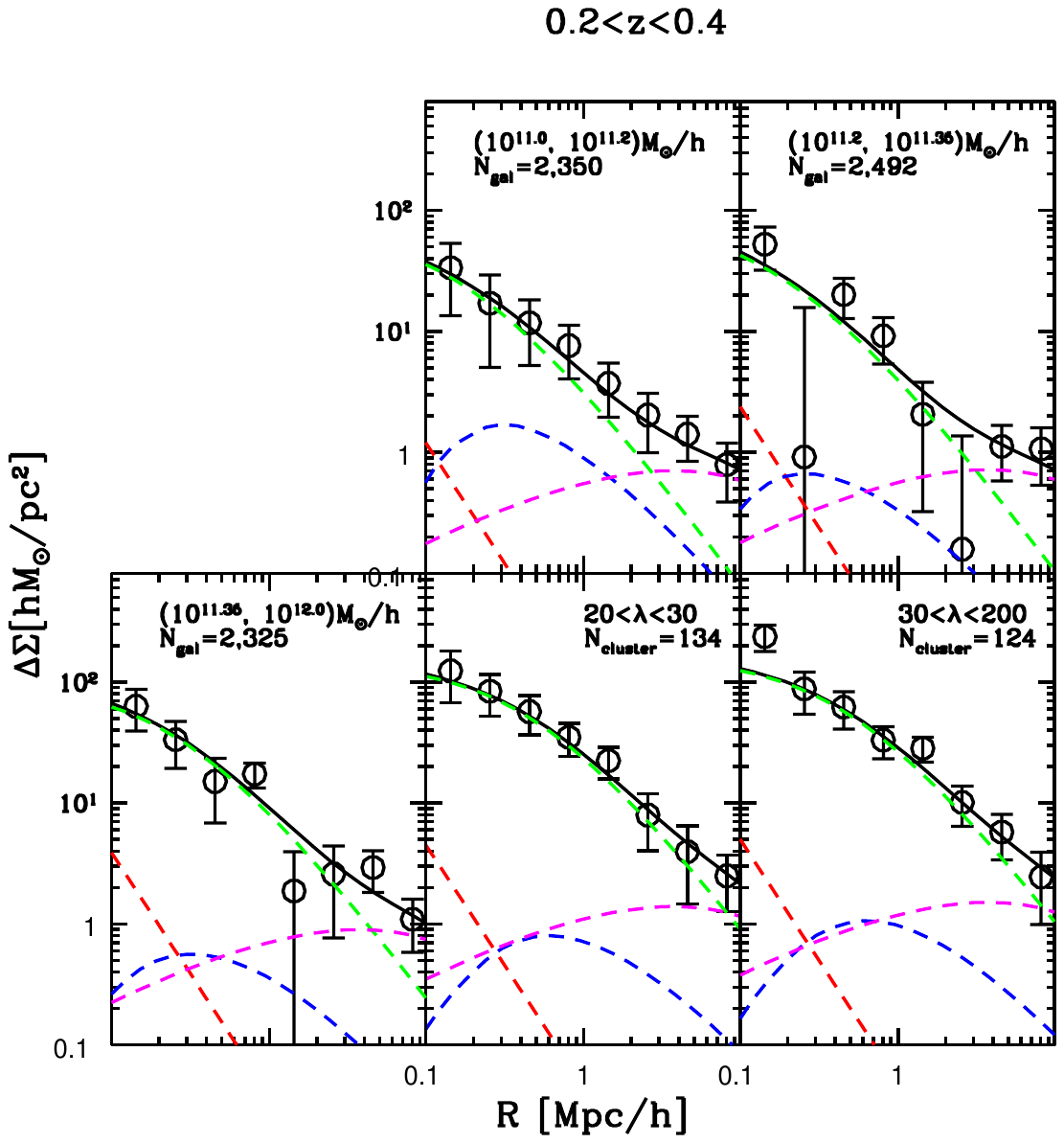}
\includegraphics[width=0.49\textwidth]{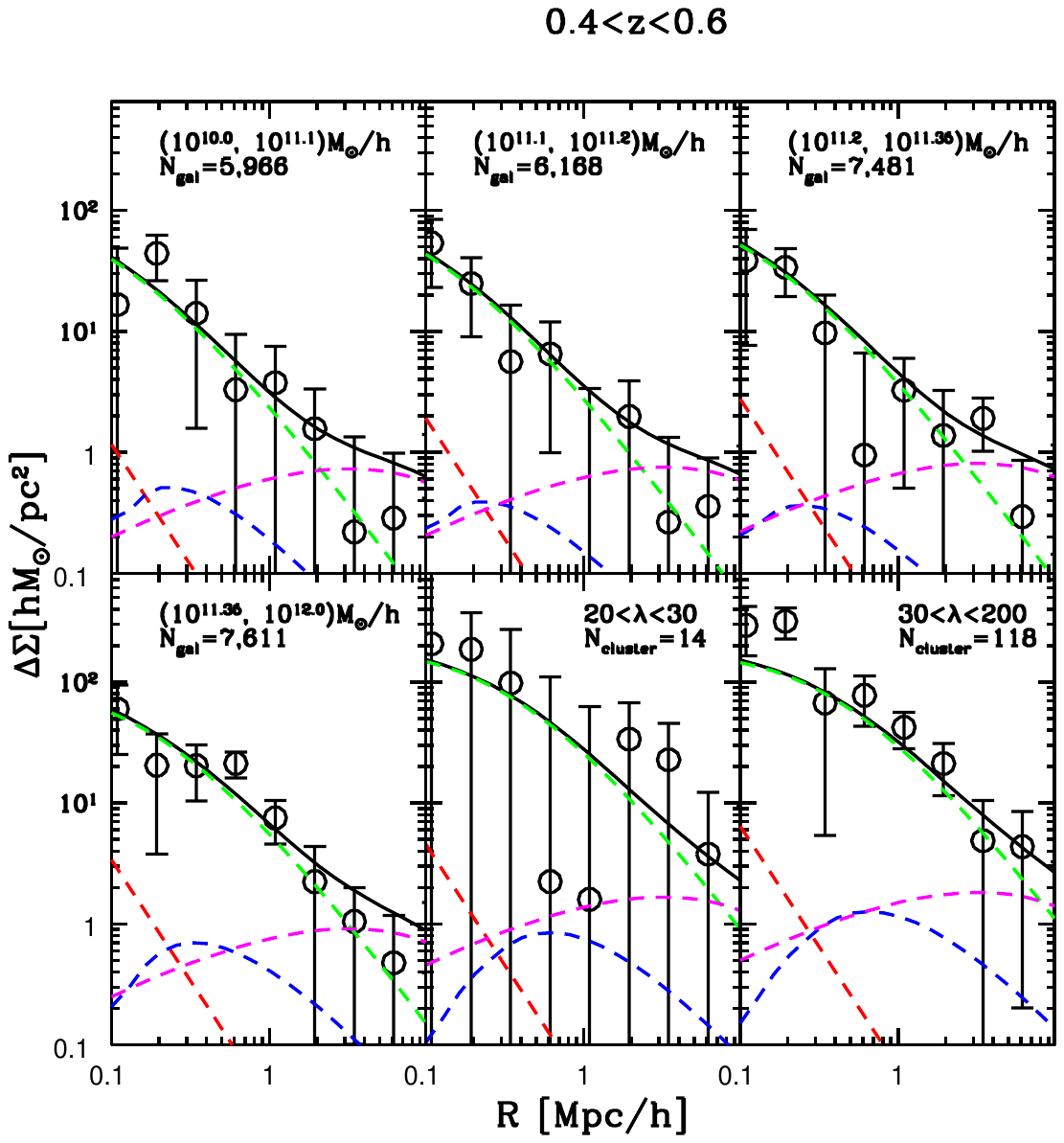}
\vspace*{-2.0cm}
\caption{Best-fit models for the low (left) and high (right) redshift samples (black solid lines). Black
circles show the measured excess surface mass density $\Delta \Sigma$ of both LOWZ/CMASS galaxies and 
clusters with different bins (clusters: Bottom-middle and bottom-right panels; LOWZ/CMASS: others). 
The model components are the central galaxy (red), the dark matter halo profile (green), the
miscentering halo component (blue) and neighboring halos (magenta).}
\label{fig:gglensing}
\end{figure*}

To interpret the average lensing signal around each sample, we fit a model including four separate components:

\vspace{1mm}

\noindent{(1) The contribution from the main dark matter halo.}

\vspace{1mm}

\noindent{(2) The contribution from the central galaxy itself; given them smallest scales to which we are sensitive are $\sim 100$ kpc, this can be treated as a point mass, as in, e.g., Leauthaud et al. (2012).}

\vspace{1mm}

\noindent{(3) A mis-centering term. In general, the position of the central galaxy is assumed to correspond to the dynamical centre of the DM halo. This may not be the case for all systems, however; simulations suggest that some principal galaxies could be ``sloshing'' around in the potential well of their halo (e.g.~Gao \& White 2006), possibly explaining features such as cold fronts in the gas in massive clusters (e.g.~Ascasibar \& Markevitch 2006). Mis-centering the halo to the primary sample will spread out and reduce the amplitude of the lensing signal on small scales, as discussed in Johnston et al. (2007), Leauthaud et al. (2010), George et al. (2012), Li et al. (2014) and  More et al. (2015).}

Following the model of Johnston et al. (2007) (subsequently tested in lensing shear maps of $25$ massive clusters by Oguri et al. 2010), the distribution of centering offsets can be described by a 2D Gaussian distribution: 
\begin{equation}
P(R_{\rm off})=\frac{R_{\rm off}}{\sigma_{\rm off}^2}\exp{\left(-\frac{1}{2}(R_{\rm off}/\sigma_{\rm off})^2 \right)},
\end{equation}
where $R_{\rm off}$ is the distance between the halo center and central galaxy position, and
$\sigma_{\rm off}$ is the effective scale length. 
We convolve this distribution of offsets with the projected density profile of the halo, to model the net contribution of mis-centering to the stacked signal. Given that the mis-centering term is subdominant in all our mass and redshift bins, we do not expect to be sensitive to this difference.

\begin{table*}
\centering
\caption{Density Profile Models in Figure~1.}
\begin{tabular}{ccccccccc}
\hline
\hline
    & z & $M_{\rm vir}$ & $c_{\rm vir}$ & $M_{200c}$ & $c_{200c}$ & $M_{*}$ & $\sigma_{\rm off}$ & $p_{\rm cen}$ \\
    &   & $10^{14} M_{\odot}/h$  &  & $10^{14} M_{\odot}/h$ & & $10^{12} M_{\odot}$ & $\rm Mpc$ & \\
\hline
$(10^{11.0}, 10^{11.2})M_{\odot}$ & $0.2<z<0.4$ & $0.13\pm0.05$  & $4.52\pm1.35$ & $0.11\pm0.04$ & $3.69\pm1.15$ & $0.13\pm0.10$  & $0.47\pm0.27$ & $0.68\pm0.15$\\
$(10^{11.2}, 10^{11.35})M_{\odot}$ & $0.2<z<0.4$ & $0.14\pm0.05$  & $4.19\pm0.81$ & $0.12\pm0.04$ & $3.41\pm0.68$ & $0.19\pm0.17$  & $0.27\pm0.18$ & $0.81\pm0.13$\\
$(10^{11.35}, 10^{12.0})M_{\odot}$ & $0.2<z<0.4$ & $0.32\pm0.07$  & $3.74\pm0.63$ & $0.27\pm0.06$ & $3.03\pm0.53$ & $0.30\pm0.27$  & $0.24\pm0.18$ & $0.85\pm0.11$\\
$20<\lambda<30$ & $0.2<z<0.4$ & $1.18\pm0.13$  & $3.34\pm1.09$ & $0.98\pm0.11$ & $2.70\pm0.91$ & $0.32\pm0.25$  & $0.61\pm0.24$ & $0.93\pm0.12$\\
$30<\lambda<200$ & $0.2<z<0.4$ & $1.44\pm0.38$  & $3.47\pm1.32$ & $1.20\pm0.32$ & $2.81\pm1.11$ & $0.37\pm0.21$  & $0.65\pm0.31$ & $0.92\pm0.10$\\
\hline
$(10^{10.0}, 10^{11.1})M_{\odot}$ & $0.4<z<0.6$ & $0.070\pm0.05$ & $5.42\pm1.39$ & $0.063\pm0.04$ & $4.63\pm1.21$ & $0.095\pm0.056$ & $0.25\pm0.19$ & $0.84\pm0.14$\\
$(10^{11.1}, 10^{11.2})M_{\odot}$ & $0.4<z<0.6$ & $0.081\pm0.05$ & $5.09\pm1.12$ & $0.072\pm0.04$ & $4.34\pm0.97$ & $0.14\pm0.11$  & $0.23\pm0.16$ & $0.87\pm0.13$\\
$(10^{11.2}, 10^{11.35})M_{\odot}$ & $0.4<z<0.6$ & $0.11\pm0.05$  & $4.82\pm0.86$ & $0.097\pm0.04$ & $4.10\pm0.75$ &$0.19\pm0.15$  & $0.23\pm0.14$ & $0.89\pm0.15$\\
$(10^{11.35}, 10^{12.0})M_{\odot}$ & $0.4<z<0.6$ & $0.18\pm0.07$  & $3.78\pm0.69$ & $0.16\pm0.06$ & $3.20\pm0.59$ & $0.30\pm0.18$  & $0.33\pm0.17$ & $0.85\pm0.22$\\
$20<\lambda<30$  & $0.4<z<0.6$ & $1.18\pm0.73$  & $3.68\pm2.66$ & $1.02\pm0.64$ & $3.11\pm2.30$ & $0.42\pm0.32$  & $0.62\pm0.51$ & $0.94\pm0.37$\\
$30<\lambda<200$ & $0.4<z<0.6$ & $1.47\pm0.28$  & $3.36\pm0.81$ & $1.27\pm0.24$ & $2.84\pm0.70$ & $0.47\pm0.28$  & $0.59\pm0.28$ & $0.91\pm0.32$\\
\hline
\hline
\end{tabular}
\leftline{Notes: $M_{*}$ is the stellar mass, derived using the Portsmouth SED-fitting code (Maraston et al. 2013), based on the SDSS-III BOSS DR10 photometry.}
\label{tab:tab1}
\end{table*}

\vspace{1mm}

\noindent{(4) Contributions from neighboring halos, which dominate the lensing profile on large scales (Johnston et al. 2007; Cacciato et al. 2009, Li et al. 2009, Oguri \& Hamana 2011). The tangential shear profile due to the neighboring halos is
\begin{equation}
\gamma_{2h}(\theta; M,z)=\int \frac{ldl}{2\pi}J_2(l\theta)\frac{\bar{\rho}_m(z)b_h(M)}{(1+z)^3\Sigma_{\rm crit}D_A^2(z)}P_m(k_l;z),
\end{equation}
where $J_2$ is the second order Bessel function, $\bar{\rho}_m(z)$ is the mass density at $z$, 
$D_A(z)$ is the angular diameter distance, $P_m(k)$ is the linear matter power spectrum, 
$b_h(M)$ is the halo bias, which we take from Tinker et al. (2010), and 
$\Sigma_{\rm crit}$ is the critical mass density
\begin{equation}
\Sigma_{\rm crit}=\frac{c^2}{4\pi G}\frac{D_s}{D_l D_{ls}},
\end{equation}
where $D_{ls}$ is the angular diameter distance between the lens and the source,
and $D_l$ and $D_s$
are the angular diameter distances from the observer to the lens and
to the source, respectively.}

\vspace{1mm}

Combining the four contributions, the lensing signal $\Delta\Sigma$ can be written as:
\begin{eqnarray}
\Delta\Sigma(R)=\frac{M_{*}}{\pi R^2}+p_{\rm cen}\Delta\Sigma_{\rm NFW}(R)+\\
(1-p_{\rm cen})\Delta\Sigma^{\rm off}_{\rm NFW}(R|P_{\rm off})+\Delta\Sigma_{2h}(R), \nonumber
\end{eqnarray}
where sequentially, the terms come from the contribution of the central galaxy, the main dark matter halo, the mis-centering effect, and the contribution of neighboring halos. 
The model is described by $5$ parameters: the halo mass $M_{\rm vir}$, the concentration $c_{\rm vir}$, 
the stellar mass $M_{*}$, the effective scale length $\sigma_{\rm off}$ and the probability $p_{\rm cen}$ that the centers we use correspond to the actual center the dark matter halos. We will fit $4$ out of the $5$ parameters in the model, and fix the stellar mass.

The stellar masses $M_*$ of LOWZ/CMASS galaxies were derived from the Portsmouth SED-fitting (Maraston et al. 2013) based on the SDSS-III BOSS DR10 data. The stellar masses of the central galaxies in the redMaPPer clusters were estimated by matching the positions of the cluster centres with the galaxy catalog of SDSS DR10.

To obtain $\Delta\Sigma$ we stack lens-source pairs in $8$ logarithmic radial ($R$) bins from $0.1$ to
$10~\rm Mpc~h^{-1}$. For a sample of selected lenses, $\Delta\Sigma(R)$ is estimated using 
\begin{equation}
\Delta\Sigma(R)=\frac{\sum_{ls}w_{ls}\gamma_t^{ls}\Sigma_{\rm crit}}{\sum_{ls}w_{ls}}\,,
\end{equation}
where $\gamma_t^{ls}$ is the tangential shear, $w_{ls}=w_n\Sigma_{\rm crit}^{-2}$, and $w_n$ is the 
\textsc{lensfit} weight factor introduced to account for intrinsic scatter in ellipticity and shape 
measurement error (Miller et al 2007, 2013). 

The lensing signal is recalibrated as in Velander et al. (2014) and Hudson et al. (2015):
\begin{equation}
\Delta\Sigma^{\rm cal}(R)=\frac{\Delta\Sigma(R)}{1+K(z_l)}.
\end{equation}
where \begin{equation}
1+K(z_l)=\frac{\sum_{ls}w_{ls}(1+m)}{\sum_{ls}w_{ls}},
\end{equation}
and the multiplicative error $m$ was determined statistically by Miller et al. (2013). 
 We apply this correction to the average shear measurement in both the low-z and high-z samples.
The effect of the correction is to increase the average lensing signals by $\sim 5.5\%$ and $6.2\%$ 
at $0.2<z<0.4$ and $0.4<z<0.6$, respectively. 

Finally, to reduce any possible `dilution' of the lensing signal by foreground galaxies scattered into our redshift selection region behind each group or galaxy, we remove any lens-source pairs with $z_s-z_l<0.1$, as well as removing any objects with a BPZ parameter $\rm ODDS<0.5$, as mentioned earlier. This corresponds to the selection {\textsc{ZCUT3}} from Battaglia et al. 2015; they show that it produces weak lensing signals comparable to their fiducial cut, and with similar errors, whereas more stringent cuts decrease the SN without any systematic change in the signal.

Considering the systematic errors in their lensing measurements due to shear calibration uncertainties and to photo-zs errors, Battaglia et al.~find that they are less than 10\%\ and 6\%\ respectively. They estimate that their total systematic uncertainties on measurements 
of $\Delta\Sigma/\Sigma_c$ are less than 16\%\ in each of their radial bins, which cover a similar range of physical or angular scales as ours, albeit stacking around slightly more massive clusters. Given that our statistical uncertainties in each bin are always more than $\sim$20\%\, and almost always more than 40\%, we will assume systematics are subdominant in our lensing measurements.

\section{Results}

We show the observed galaxy-galaxy lensing signals for the low and high redshift samples in Figure~1.
The black circles represent the measured excess surface mass density. 
Errorbars shown are 1$\sigma$ fluctuations obtained using $1000$ bootstrap by resamplings $30$ equal sub-areas 
from the real observation data sets.

We fit the measured lensing signals to the model described in section~3 using a Monte Carlo Markov Chain (MCMC) method. A flat prior is adopted for the halo mass $M_{\rm vir}$, the concentration $c_{\rm vir}$ and the probability 
$p_{\rm cen}$. A Gaussian prior is adopted for the
effective scale length $\sigma_{\rm off}$ as Johnston et al. (2007). 

As above, we select the clusters and massive galaxies carefully. For the cluster, only clusters with 
well-defined centers with $p_{\rm cen}>0.9$ are selected; for the LOWZ/CMASS galaxies, they are almost
all central galaxies, $\sim10\%$ of them may be the most luminous satellite galaxies as indicated by the clustering
analysis by Parejko et al. (2013) and White et al. (2011). Then a flat prior $(0.65, 1.0)$ on the probability
$p_{\rm }cen$ is adopted. A Gaussian prior is adopted for the effective scale length $\sigma_{\rm off}$ as 
Johnston et al. (2007). Moreover, We test the effects of the adopted prior with a Gaussian one used in 
Melchior et al. (2016) for the $p_{\rm cen}$, the difference on the fit results of mass and concentration 
is $<5\%$, which suggests that such an effect is sub-dominant.
 
The black solid curves shows the sum of all components for the best-fit models. The different model components are shown as dashed curves: the central galaxy (red), the dark matter halo profile (green), the mis-centering component (blue), and the contribution from neighboring halos (magenta).  At small scales ($0.1-1 \rm Mpc$), the dark matter halo is the dominant component. At large scales ($>3 \rm Mpc$ for galaxies and $>5 \rm Mpc$ for clusters), the contribution from neighboring halos dominates. The best-fit results for the 5 parameter values are given in Table~1. We convert the measured $M_{\rm vir}$ and $c_{\rm vir}$ to $M_{200}$ and $c_{200}$ with 
the formula in Johnston et al. (2007). 
 
The values of the fitted probability $p_{\rm cen}$ were all higher than $0.9$, which indicates that: (1) the centers of redMaPPer 
clusters are well defined; (2) the centers of $\sim 10\%$ LOWZ/CMASS galaxies are offset.

As Cibirka et al. (2017), we investigate the impact of the intrinsic scatter of mass-richness relation 
for the clusters with simulations. With the richness of redMaPPer clusters, we can estimate the related mass 
from the mass-richness relation with $\sigma_{m|\lambda}=0.25\rm dex$ (Simet et al. 2017), and draw the related
concentraion from \textit{m-c} relation with $0.1\rm dex$ (Dutton \& Maccio 2014). We can then generate the 
$\Delta \Sigma$ for each cluster with the related mass and concentration. With $1000$ independent analysis
for the four redMaPPer cluster bins using the same analysis procedure, we found that the difference between 
the mean input and fitted average mass and concentration are $\sim0.025\pm0.020$ and $\sim0.23\pm0.31$, 
respectively. We conclude that it can be negligible in comparison with the measurement uncertainties.

\subsection{\textit{m-c} relation}

The recent Multidark simulations of Klypin et al. (2014) predict a \textit{m-c} relation of the form:
\begin{equation}
c_{200c}(z,M)=A\left(\frac{M_{200c}}{M_0}\right)^{B}(1+z)^{C},
\end{equation}
where we will take $M_0=2.0 \times 10^{12}M_{\odot}h^{-1}$. 
Here $M_{200c}$ and $c_{200c}$ denote the mass and concentration with respect to $R_{200c}$, the radius within which the halo has 200 times the critical density of the universe at that redshift.

As we cannot test the predicted redshift dependence of the \textit{m-c} relation very accurately with only
two redshift bins, we will assume the redshift dependence predicted by Klypin et al..
For the redshift range $0.2<z<0.6$ and the mass range of our samples, this corresponds to an exponent $\rm C\sim-0.67$.
The best-fit \textit{m-c} relations are shown in Table~2. We note that in practice, given the errors on the slope and normalization for the low-redshift sample, the \textit{m-c} relations of both samples are actually consistent with no redshift evolution. 

Giocoli et al. (2012) found that halo triaxiality and substructures within the host halo virial radius can bias the observed, 2D
\textit{m-c} relation. They propose a method for correcting the observed 2D \textit{m-c} relation for projection effects:
\begin{equation}
c_{2D}(M)=c_{3D}(M)\times 1.630M^{-0.018},
\end{equation}
The exact magnitude of the correction depends on the the radial range over which the fit is performed. 

Meneghetti et al. (2014) studied the projection effects on the concentration with MUSIC-2 halos and found 
that the trend is qualitative agreement with the results of Giocoli et al. (2012). The correction 
by Giocoli et al. (2012) is obtained with the simulated halos of mass of $5\times10^{13}M_{\odot}/h$ for 
$z<0.4$ and $1.5\times10^{14}M_{\odot}/h$ for $z>0.4$, which is roughly consistent with the cluster 
mass region of our data, we can apply the correction as listed above to correct the 2D results of cluster 
measurement back to 3D results. However, there are no correction estimation for the lower mass DM halo. 
In order to be consistent, we extrapolate the correction to the lower halo mass region for the LOWZ/CMASS 
galaxies in this analysis.

In Figure~2, we show the resulting 3D, corrected \textit{m-c} relation, after rescaling to $z=0$. The 3D correction is used on all the observational dataset predictions.

\begin{table}
\centering
\caption{Best-fit Mass-Concentration relations.}
\begin{tabular}{cccc}
\hline
\hline
 z   & A & B \\
\hline
$0.2<z<0.4$ & $5.24\pm1.24$ & $-0.13\pm0.10$ \\
$0.4<z<0.6$ & $6.61\pm0.75$ & $-0.15\pm0.05$ \\
\hline
\hline
\end{tabular}
\label{tab:tab2}
\end{table}

\begin{figure}
\includegraphics[width=0.5\textwidth]{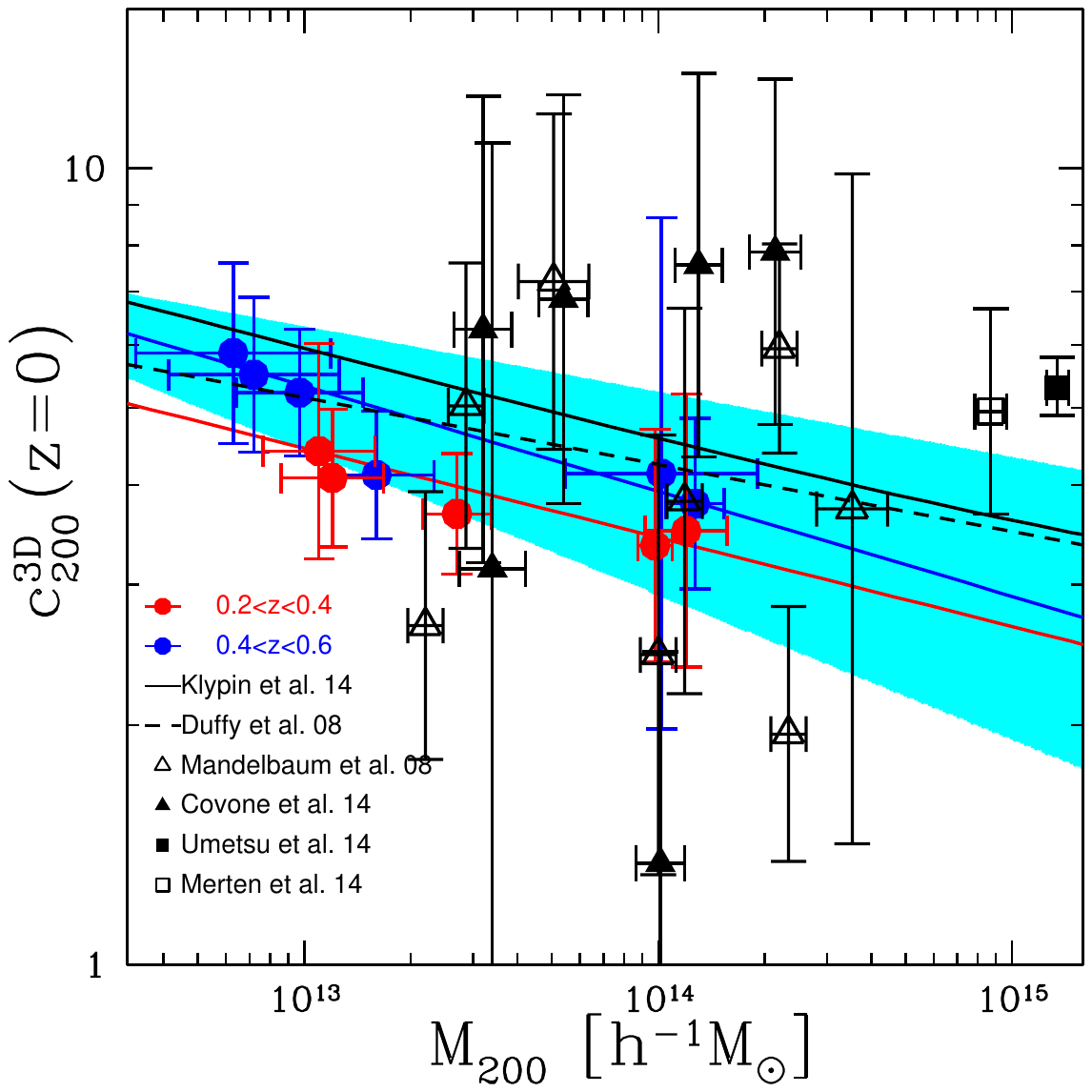}
\vspace*{-2.2cm}
\caption{The 3D corrected \textit{m-c} relation for galaxies and clusters in the CS82 survey, after rescaling all the results to $z=0$, assuming the redshift evolution from Klypin et al. (2014). Red and blue circles denote the stacked lensing signals from low and high redshift samples, respectively. The red curve is the best-fit \textit{m-c} relation for the low redshift sample. The blue curve and cyan-shaded area is best-fit \textit{m-c} relation for the high redshift sample and its $1\sigma$ uncertainty. Black curves are the simulation predictions by Duffy et al. (2008) (dashed) and Klypin et al. (2014) (solid). The black symbols denote the lensing-based measurements of concentration and mass by Mandelbaum et al. (2008) with SDSS (median redshift $\bar z\sim 0.22$); Covone et al. (2014) with CFHTLenS ($\bar z\sim 0.36$); and Umetsu et al. (2014) ($\bar z\sim 0.35$) and Merten et al. (2014) $\bar z\sim 0.40$ with the CLASH cluster sample. For clarity, we have binned the data points of Merten et al. (2014). Note that in this paper the overdensity parameter for all mass definitions is relative to critical density $\rho_{crit}$.}
\label{fig:mc}
\end{figure}

Given the corrections from 2D to 3D by Giocoli et al. (2012), we can directly compare our measurements with the 3D simulation results. Comparing with the results by Duffy et al. (2008) and Klypin et al. (2014), the amplitudes $\rm A$ and slopes $\rm B$ of both 
redshift samples are consistent with the simulation predictions. 

We can also compare our results with previous lensing measurements. As with our own results, we correct all
observational measurements from 2D to 3D using the formula of  Giocoli et al. (2012). 

By stacking weak lensing signals of galaxies, groups and clusters from SDSS dataset, Mandelbaum et al. (2008) 
fit the data to the model assuming a spherical NFW profile excluding small scales to reduce the effects of 
baryons and miscentering halos. As shown in Figure~2, their measurements are comparable to ours, but are 
more scattered. 

Using photo-z selected clusters (Wen et al. 2009) in the CFHTLenS data, Covone et al. (2014) found 
a \textit{m-c} relation with a normalization 1$\sigma$ above that of Duffy et al (2008), but were not able to measure the slope very accurately.

The recent weak+strong lensing measurements for individual clusters in the CLASH sample (Merten et al. 2014; shown binned here), as well as stacked shear+magnification measurements for the same sample (Umetsu et al. 2014), probe the concentration of more massive clusters with $M_{200} \sim 10^{15}M_{\odot}h^{-1}$. Generally, they find mean concentrations 1--2$\sigma$ above our best-fit relations at lower mass. This could indicate a change in the slope of the \textit{m-c} relation at high mass, but it may also reflect the properties of the sample. Using hydrodynamical simulations, Meneghetti et al. (2014) have shown that the CLASH selection function, 
based on degree of relaxation as estimated from X-ray morphology, biases the sample to higher concentrations, such that a higher normalization and a steeper slope to the \textit{m-c} relation is expected. Strong-lensing selected samples are also expected to behave similarly.

Another possible interpretation of the higher concentrations in massive systems, however, is that they are real, and reflect a systematic change in the density profile, from NFW-like to a different form that has a more concentrated mass distribution for a given value of the scale and virial radii. As noted previously, high resolution simulations suggest that the most massive clusters, corresponding to high peaks in the initial density field, should have density profiles that differ slightly from the NFW form, and are better represented by the Einasto form (e.g.~Gao 2008; Klypin et al 2014). For typical values of the Einasto shape parameter $\alpha$, the density profile shows more curvature in its logarithmic slope than an NFW profile.

The halos in our sample correspond to peaks of height $\nu \equiv \sigma(M)/\delta_c\simeq1.0$--2.7. Using the scaling relation from Gao et al. (2008), we expect an Einasto parameter $\alpha$ in the range 0.165 to 0.225. The corresponding 3D density profiles are very close to the original NFW form. Over the range where the main halo term should dominate our lensing measurements, $\sim 0.25 r_s$ (for the most massive, low concentration, low redshift clusters) to 2--3 $r_{vir}$, Einasto profiles with these values of $\alpha$ differ from the NFW form by less than 20\%, and this difference is largest at several times $r_{vir}$. Over the range 0.25 $r_s$ -- 2 $r_{vir}$, the 3D profiles differ from an NFW profile by less than 7\%. Klypin et al.~predict a slightly larger range of $\alpha$ for our peak heights, but here too the final 3D profiles differ from NFW by less than 20\%. Projection, halo-to-halo scatter, and the other contributions to the surface density profile will further reduce the difference between the two models, so we do not expect to be sensitive to this difference. We have tested fitting the lensing signals of the redMaPPer clusters with the Einasto profile, and find a similar \textit{m-c} relation as for the NFW profile. We conclude that our data can not distinguish between the two profiles on the mass scales we probe, as expected. Measurements at higher halo mass should be better able to confirm the predicted variation in the shape of the profile (although even with the much more massive CLASH sample, Umetsu et al. (2014) currently find a mean profile consistent with the original NFW form, and no improvement in the fit when using the Einasto form).

In this paper, the redMaPPer clusters are stacked by richness, while the LOWZ/CMASS galaxy are
stacked by stellar mass.

For the cluster, a CLASH X-ray selected sample can present a bias of $\sim10\%$ on concentration 
(Meneghetti et al. 2014). Because the redMaPPer cluster sample in our analysis is more general and less 
constraining, we expect the effects is at the few percent level. Furthermore, Old et al. (2015) found that no 
systematic bias exists for the cluster mass estimation with the redMaPPer algorithm, which analyze the mass 
determination using the abundance matching method. The impact of selection effect in the \textit{m-c} relation 
can be ignored.

For the galaxy, using the EAGLE simulation, Matthee et al. (2016) studied the the magnitude and origin 
of the scatter in the \textit{SHMR}. They found that a larger stellar mass corresponds to a more concentrated
halo at fixed halo mass. However, the inclusion of concentration does not affect the scatter in
stellar mass for the halo mass $M_{200}>10^{12.5}~\rm M_{\odot}$. Because the lowest DM halo mass of our
measurement is $>5\times10^{12}~\rm M_{\odot}/h$, the bias on the \textit{m-c} relation from the
stellar mass scatter is insignificant.

\begin{table*}
\centering
\caption{Best-fit results for the \textit{SHMR}.}
\begin{tabular}{cccccc}
\hline
\hline
z & $\log_{10}(M_1)$ & $\log_{10}(M_{*,0})$ & $\beta$ & $\delta$ & $\gamma$ \\
\hline
$0.2<z<0.4$ & $12.52 \pm 0.050$ & $10.98\pm0.036$ & $0.47\pm0.022$ & $0.55\pm0.13$ & $1.43\pm0.28$\\
$0.4<z<0.6$ & $12.70 \pm 0.057$ & $11.11\pm0.038$ & $0.50\pm0.025$ & $0.54\pm0.16$ & $1.72\pm0.30$\\
\hline
\hline
\end{tabular}
\label{tab:tab3}
\end{table*}

\begin{figure}
\includegraphics[width=0.5\textwidth]{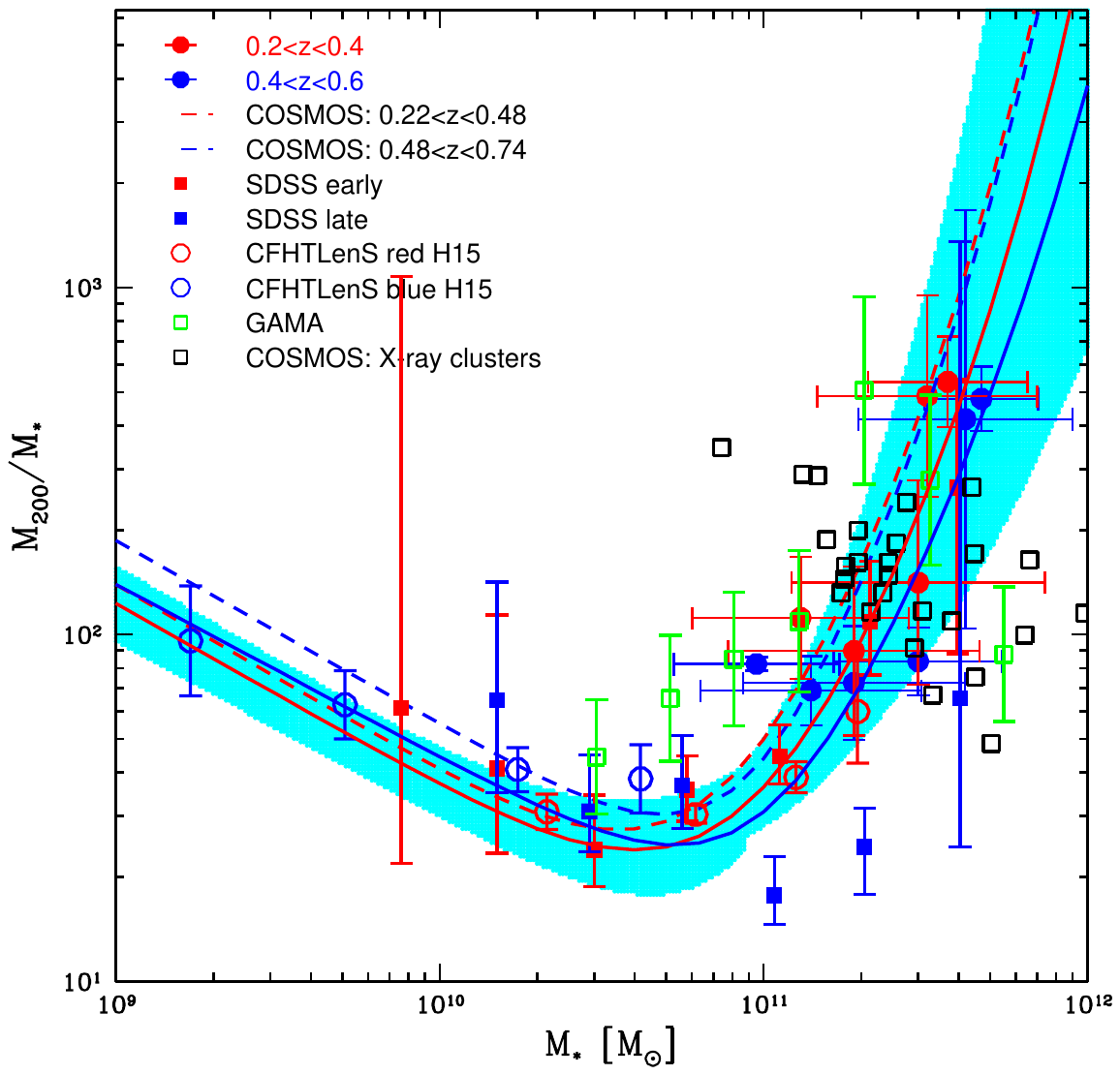}
\vspace*{-2.2cm}
\caption{Halo-to-stellar mass ratio (the inverse of the SHMR) as a function of stellar mass.
Red and blue dots denote the measurements for the low and high redshift samples,
respectively. The red solid curve and cyan-shaded area is the best-fit \textit{SHMR}
relation and its $1\sigma$ uncertainty derived by fitting our low redshift sample combined with the results of H15. 
The blue solid curve is the best-fit \textit{SHMR} relation of our high redshift sample combined with H15.
The red and blue dashed curves are the relations fitted by Leauthaud et al. 2012, using COSMOS measurements
We also compare our measurements with other galaxy-galaxy lensing measurements, including 
SDSS (Mandelbaum et al. 2006), CFHTLenS (H15), COSMOS X-ray clusters (Leauthaud et al. 2010; 2012), and GAMA (Han et al. 2015).}
\label{fig:ms}
\end{figure}

\subsection{\textit{SHMR} measurement}

Given a mean stellar mass for the central galaxies in each sample, we can also estimate the \textit{SHMR}. 
Following Behroozi et al. (2010) and Leauthaud et al. (2012), the SHMR is modelled as a log-normal probability 
distribution function, with a mean log relation denoted as $M_*=f_{SHMR}(M_h)$, and a (log-normal) scatter denoted $\sigma_{\log M_*}$.

The relation $f_{SHMR}(M_h)$ is mathematically defined via its inverse 
function:
\begin{eqnarray}
\log_{10}(f^{-1}_{SHMR}(M_h))=\log_{10}(M_1)+\beta \log_{10}\left(\frac{M_*}{M_{*,0}}\right)+ \\
\frac{\left(\frac{M_*}{M_{*,0}}\right)^{\delta}}{1+\left(\frac{M_*}{M_{*,0}}\right)^{-\gamma}}-\frac{1}{2}, \nonumber
\end{eqnarray}
where $M_1$ is a characteristic DM halo mass, $M_{*,0}$ is a characteristic stellar mass, 
$\beta$ is the low-mass \textit{SHMR} slope, and $\delta$ and $\gamma$ jointly determine the high-mass \textit{SHMR} slope (Behroozi et al. 2010). 

Since our data only probes the high-mass end of this relation, we combine our results with those of Hudson et al. 2015 (H15), which cover the same redshift bins as ours but sample much smaller masses, in order to fit the full \textit{SHMR}.
We note that the SHMR is defined to have log-normal scatter in $M_*$ at fixed $M_h$, whereas we are binning data in bins of fixed $M_*$. As a result, we need to fit to the mean of the inverse relation, which because of the non-linearity and scatter is not equal to the inverse of the mean relation. We correct the mean halo mass at a given stellar mass for the effect of the scatter $\sigma_{\log M_*}$ by including a correction factor derived using the same method as Velander et al. (2014). We adopt a scatter of $\sigma_{\log M_*} = 0.2\rm dex$ in our analysis as Leauthaud et al. (2012).
Given this correction, the results of our fits for the high and low redshift datasets, combined in each case with H15,  are given in Table~3.

As discussed in Leauthaud et al.~(2012) and Behroozi et al.~(2010), the general shape of the SHMR (or its inverse, shown in Figure~3) indicates a peak in the efficiency of star formation for galaxies with masses around $M_*\sim$4--6$\times10^{10}M_{\odot}$. Below this, stellar feedback limits the growth of galaxies, while above this mass hot gas and AGN are probably responsible for the decrease in the stellar fraction. In Figure~3, we show the binned results and best fit relations, in addition to results from other galaxy-galaxy lensing studies, including the measurements in SDSS (Mandelbaum et al. 2006), CFHTLenS (H15), 
the Cosmic Evolution Survey (COSMOS: Leauthaud et al. 2010, 2012) and the Galaxy And Mass Assembly (GAMA: 
Han et al. 2015). In general there is good agreement between the different lensing determinations, as well as with other techniques such as abundance matching or clustering (e.g.~Behroozi et al.~2013; Rodriguez-Torres et al.~2015). The only possible disagreement is at large masses, where the scatter in individual measurements is large. The individual X-ray clusters studied by Leauthaud et al. (2010; 2012), for instance, differ from the mean relation by factors of up to 5 (or 0.7 dex), which seems unlikely to be consistent with our assumed scatter of $\sigma_{\log M_*}=0.2\,\rm dex$, given the small size of that sample. 

%Mandelbaum et al. (2016) measured the mean halo mass as a function of the stellar mass with a sample of 
%Locally Brightest Galaxies (LBG) from the SDSS. The SHMR measured with the red sample is inconsistent with 
%our results.

This scatter for individual systems suggests that selection effects may still be a concern, in our and other published determinations of the SHMR. For instance, Leauthaud et al. (2015) found that the CMASS sample is significantly affected by mass incompleteness, and is $\sim 80~\%$ complete at $\log_{10}(M_{*}/M_{\odot})>11.6$ only in the narrow redshift range $z=[0.51, 0.61]$. The LOWZ sample, on the other hand, is $\sim 80~\%$ complete at $\log_{10}(M_{*}/M_{\odot})>11.6$ for $z=[0.15, 0.43]$. If completeness is correlated with high or low stellar mass for the primary galaxy, it could mean that the true scatter is larger than we have assumed. Follow-up studies of the individual objects in our samples, as well as data from larger forthcoming lensing surveys, will be required to determine the exact effect of these biases on the mean relation.

A Chabrier (2003) IMF was assumed by the measurements of H15, Leauthaud et al. (2012) and Han et al. (2015).
Ilbert et al. (2010) investigated the possible sources of uncertainty and bias by comparing stellar mass
estimates between methods. The expected difference between a Kroupa IMF and a Chabrier IMF is not significant.

\section{Conclusions}

Combining samples of massive galaxies and clusters, from the redMaPPer and LOWZ/CMASS catalogues respectively, with the shear catalog of the CS82 survey, we have measured the projected density profiles around central galaxies at two different redshifts.
 We fit the lensing shear signal around these galaxies as a sum of contributions from the main dark matter halo, the central galaxy, centering errors and physical offsets between the galaxy and the halo, and neighboring halos. From the resulting fits, we determine 
 the mean halo concentration as a function of mass, and also the mean relation between (central) stellar mass and halo mass (the \textit{SHMR}).

\noindent{Our main results are as follows:}

\vspace{1mm}

\noindent{1. We find that halo concentration decreases weakly with mass, $c=A(M/M_0)^{B}$, with the amplitude ${\rm A} \sim$5--6 and
slope $\rm B\sim -0.14$. These parameter values are in good agreement with the predictions of simulations, e.g. Duffy et al. 2008 or Klypin et al. 2014, given a correction for projection (Giocoli et al. 2012). Our low normalization also suggests that the much higher values $c \gtrsim 10$ measured for more massive cluster samples are indeed a result of selection and/or projection effects, as inferred previously, e.g.~by Meneghetti et al.~(2014) for the CLASH sample.}

\vspace{1mm}

\noindent{2. Comparing our halo masses to the stelar masses of the central galaxies from the SDSS DR10 catalogues, we derive the SHMR for massive systems at redshift 0.2--0.6. Combining these measurements with those of H15 at lower masses, we fit the broken power-law form proposed by Behroozi et al.~(2010), and find results consistent with those of previous lensing studies, e.g.~Leauthaud et al.~(2012).}

\vspace{1mm}

In general, there is reasonable agreement between many different surveys and techniques on the overall shape of the \textit{SHMR}. One unresolved issue is the extent of the scatter, particularly at the high-mass end. In groups and clusters, there could be a number of reasons for this scatter, including variations in halo merger history, hot gas content or thermodynamic state, AGN activity, or the details of central galaxy mergers. Having calibrated the mean relation for the redMaPPer and LOWZ/CMASS samples in CS82, we can follow-up these possibilities with the ancillary data available in this field. Future wide-field lensing surveys will also provide far more sensitive lensing measurements, sufficient to determine masses and concentrations for restricted subsets of groups and clusters, and to test for correlation between residuals in the halo scaling relations, and the detailed properties of individual galaxies, groups and clusters.

\section*{Acknowledgments}
This work is based on observations obtained with MegaPrime/MegaCam, a joint project of CFHT and CEA/DAPNIA, at the
Canada-France-Hawaii Telescope (CFHT), which is operated by the National Research Council (NRC) of Canada,
the Institut National des Science de l'Univers of the Centre National de la Recherche Scientifique (CNRS) of
France, and the University of Hawaii. The Brazilian partnership on CFHT is managed by the Laborat\'{o}rio Nacional
de Astrof\'isica (LNA). This work made use of the CHE cluster, managed and funded by ICRA/CBPF/MCTI, with financial
support from FINEP and FAPERJ. We thank the support of the Laborat\'{o}rio Interinstitucional de e-Astronomia (LIneA).
We thank the CFHTLenS team for their pipeline development and verification upon which much of this surveys pipeline was
built.

The authors thank Francisco Prada and Charling Tao for useful discussions.
This research was supported by a Marie Curie International Incoming
Fellowship within the $7^{th}$ European Community Framework Programme. 
HYS acknowledges the support from Swiss National Science Foundation (SNSF) and NSFC of China under grants 11103011. 
JPK acknowledges support from the ERC advanced grand LIDA. 
LR acknowledges the NSFC grants (No.11303033) and the support from Youth Innovation Promotion Association of CAS.
JC acknowledges financial support from MINECO (Spain) under project number AYA2012-31101. 
TE is supported by the Deutsche Forschungsgemeinschaft through the Transregional Collaborative Research Centre TR 33 - 'The Dark Universe'. 
MM is partially supported by CNPq (grant 486586/2013-8) and FAPERJ (grant E-26/110.516/2-2012).
BM acknowledges financial support from the CAPES Foundation grant 12174-13-0.
JET acknowledges support from  a Discovery Grant from the Natural Sciences and Engineering Research Council (NSERC) of Canada.

\end{document}